\documentclass[useAMS,usenatbib,usegraphicx]{mn2e}

\newcommand{\microns}{$\mu$m}

\def\ga{\mathrel{\hbox{\rlap{\hbox{\lower4pt\hbox{$\sim$}}}\hbox{$>$}}}}
\def\la{\mathrel{\hbox{\rlap{\hbox{\lower4pt\hbox{$\sim$}}}\hbox{$<$}}}}

\def\msun{$M$\mbox{$_{\normalsize\odot}$}}

\def\vlsr{$v_{\rm LSR}$}

\def\kms{\,km~s$^{-1}$}

\def\arcsec{$^{\prime \prime}$}

\newcommand{\fig}[1]{Fig.\ \ref{#1}}

\title[The Massive Intermediate-age Star Cluster GLIMPSE-CO1]{GLIMPSE-CO1:
  the most massive intermediate-age stellar cluster in the Galaxy} 
\author[Davies et al.]{Ben Davies$^{1,2}$, Nate Bastian$^{3,4}$, Mark
  Gieles$^{4}$, Anil C.\ Seth$^{5}$, Sabine Mengel$^{6}$, and \newauthor Iraklis S.\ Konstantopoulos$^{7}$
  \\ $^{1}$Rochester Institute of Technology, 54 Lomb Memorial Drive,
  Rochester, NY 14623, USA\\ $^{2}$School of Physics \& Astronomy,
  University of Leeds, Woodhouse Lane, Leeds LS2 9JT,
  UK\\
 $^{3}$School of Physics, University of Exeter, Stocker Road, Exeter
  EX4 4QL, UK \\ 
 $^{4}$Institute of Astronomy, University of
  Cambridge, Madingley Road, Cambridge CB3 0HA, UK\\
 $^{5}$Harvard-Smithsonian Center for Astrophysics, 60 Garden Street,
Cambridge MA 02138, USA \\
 $^{6}$European Southern Observatory, Karl-Schwarzschild-Str. 2, 85748 Garching, 
Germany \\
 $^{7}$Eberly College of Science, The Pennsylvania State University, 525
Davey Lab, University Park, PA 16802, USA \\
}
\begin{document}

\date{}

\pagerange{\pageref{firstpage}--\pageref{lastpage}} \pubyear{2010}

\maketitle

\label{firstpage}

\begin{abstract}
The stellar cluster GLIMPSE-C01 is a dense stellar system located in
the Galactic Plane. Though often referred to in the literature as an
old globular cluster traversing the Galactic disk, previous
observations do not rule out that it is an intermediate age (less than
a few Gyr) disk-borne cluster. Here, we present high-resolution
near-infrared spectroscopy of over 50 stars in the cluster. We find an
average radial velocity is consistent with being part of the disk, and
determine the cluster's dynamical mass to be $(8 \pm 3)
\times$10$^4$\msun. Analysis of the cluster's $M/L$ ratio, the
location of the Red Clump, and an extremely high stellar density, all
suggest an age of 400-800Myr for GLIMPSE-C01, much lower than for a
typical globular cluster. This evidence therefore leads us to conclude
that GLIMPSE-C01 is part of the disk population, and is the most
massive Galactic intermediate-age cluster discovered to date.
%, and therefore that it is part of the disk population,
%though decisive evidence remains elusive. We suggest future
%observational tests which should provide a definitive answer as to the
%nature of GLIMPSE-C01.
\end{abstract}


\section{Introduction} \label{sec:intro}
With the advent of large near/mid-infrared (IR) detectors and
wide-area surveys, many massive stellar clusters ($>10^{4}$\msun) have
recently been discovered in the plane of the Galaxy. So far, due to
the selection techniques, only extremely young clusters ($<$ 25 Myr)
have been identifed
\citep[e.g.][]{Clark05,Figer06,RSGC2paper}. However, based on
size-of-sample effects \citep{G-B08} and what we see in similar face
on spiral galaxies \citep{L-R99} we expect a large number of
intermediate age (100-1000 Myr) clusters with masses between
(10$^5$-10$^6$\msun). Due to their stellar population properties,
these clusters are expected to be very difficult to detect due to
confusion with background stars.

A candidate for such an intermediate age cluster may be GLIMPSE-C01
(hereafter GC01), an object discovered in the Spitzer GLIMPSE mid-IR
survey of of the Galactic disk \citep[][hereafter
  K05]{Kobulnicky05}. Despite the cluster being located 0.1 degrees
from the Galactic plane, these authors suggest that G01 is an old
globular cluster which happens to be passing through the disk of the
galaxy. However, they note that the colour-magnitude diagram (CMD) is
equally well fit by a intermediate age stellar population ($<$ few
Gyr; see Fig. 7 in K05), while it may be argued that the cluster's
luminosity function provided inconclusive results (their Fig.\ 8).
Further studies of the cluster have not provided convincing evidence
of the cluster's age or metallicity. \citet{Ivanov05} provided higher
resolution near-IR photometry, detecting what appears to be the
Red-Clump in the colour-magnitude diagram (see their Fig.\ 4), and
argued from the slope of the Red Giant Branch (RGB) that the cluster
was metal poor. However, as stated by Ivanov et al., their method
makes the implicit assumption that the cluster was $\ga$10Gyr old, and
is not calibrated for intermediate-age clusters.

\begin{figure}
  \centering
  \includegraphics[width=8cm]{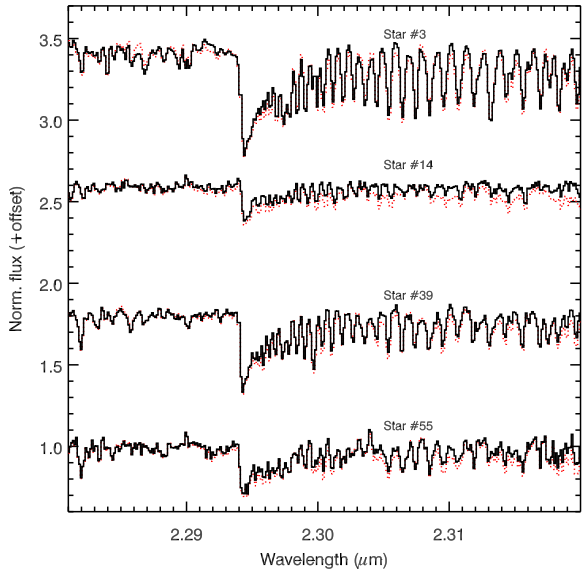}
  \caption{Examples of the data obtained, while in red we show the
    spectra prior to the fringing correction. In order from top to
    bottom, we show a bright object ($K$=8.31) with strong CO; a
    bright object ($K$=9.68) with weak CO; a faint object ($K$=10.87)
    with strong CO; and finally the spectrum of the faintest star we
    obtained ($K$=12.13).  }
  \label{fig:showspec}
\end{figure}

In this paper we present high resolution near-IR spectroscopy of
several stars in the cluster. From these data we are able to determine
radial velocities for each star observed, measure the cluster's
velocity dispersion, and ultimately the cluster's dynamical
mass. Additionally, we obtain a very reliable measurement of the
cluster's average radial velocity, which we can compare to the
Galactic velocity field to establish whether or not the cluster is
co-moving with the disk. We combine this new information with that
already present in the literature to attempt to shed new light on the
nature of GC01 -- is it an old globular cluster passing through the
disk, or is it an intermediate age disk-borne cluster, the first of
it's kind detected in our Galaxy?

We begin in Sect.\ \ref{sec:obs} with a description of the
observations, data reduction procedures and analysis techniques. In
Sect.\ \ref{sec:results} we describe our results, and discuss the
issues of the cluster's physical properties such as distance and
mass. The nature of GC01 is discussed in Sect.\ \ref{sec:disc}, while
we conclude in Sect.\ \ref{sec:conc}.

\section{Observations \& data reduction} \label{sec:obs}
Our data were taken as part of the ESO observing programme 383.D-0025
(PI N.\ Bastian). We obtained near-infrared spectra of numerous stars
in the field of GCO1 using ISAAC \citep{Moorwood98}, mounted on UT1 of
the ESO-VLT. We used the instrument in medium resolution mode with the
0.3\arcsec\ long-slit and central wavelength of 2.3\microns. This
set-up achieves a spectral resolution of $R$=8900 in the range
$\sim$2.25--2.35\microns, allowing us to observe the CO bandhead at
2.293\microns\ and neighbouring continuum.

\begin{figure*}
  \includegraphics[width=8.5cm,bb=0 0 623 433]{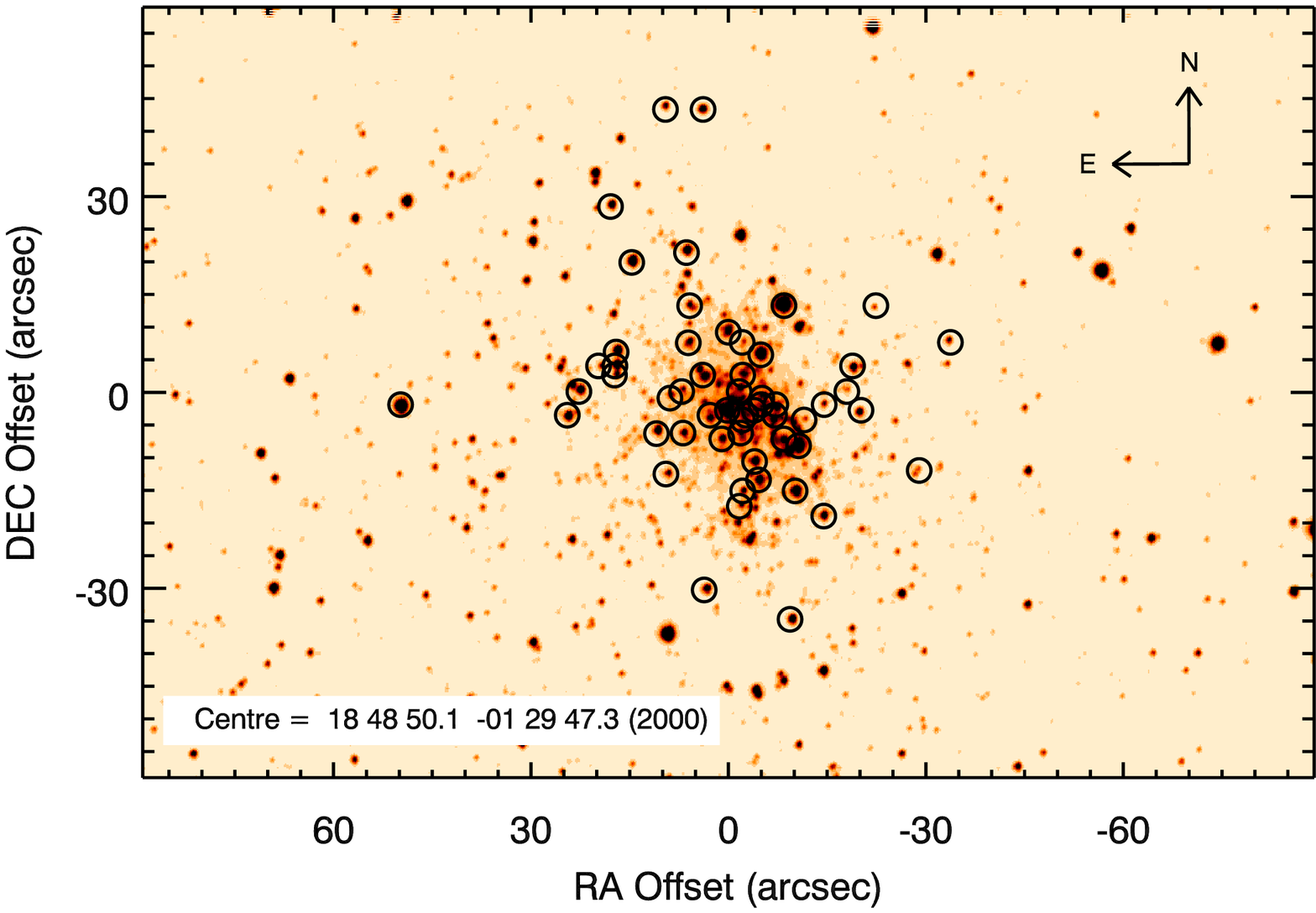}
  \includegraphics[width=8.5cm,bb=0 0 623 433]{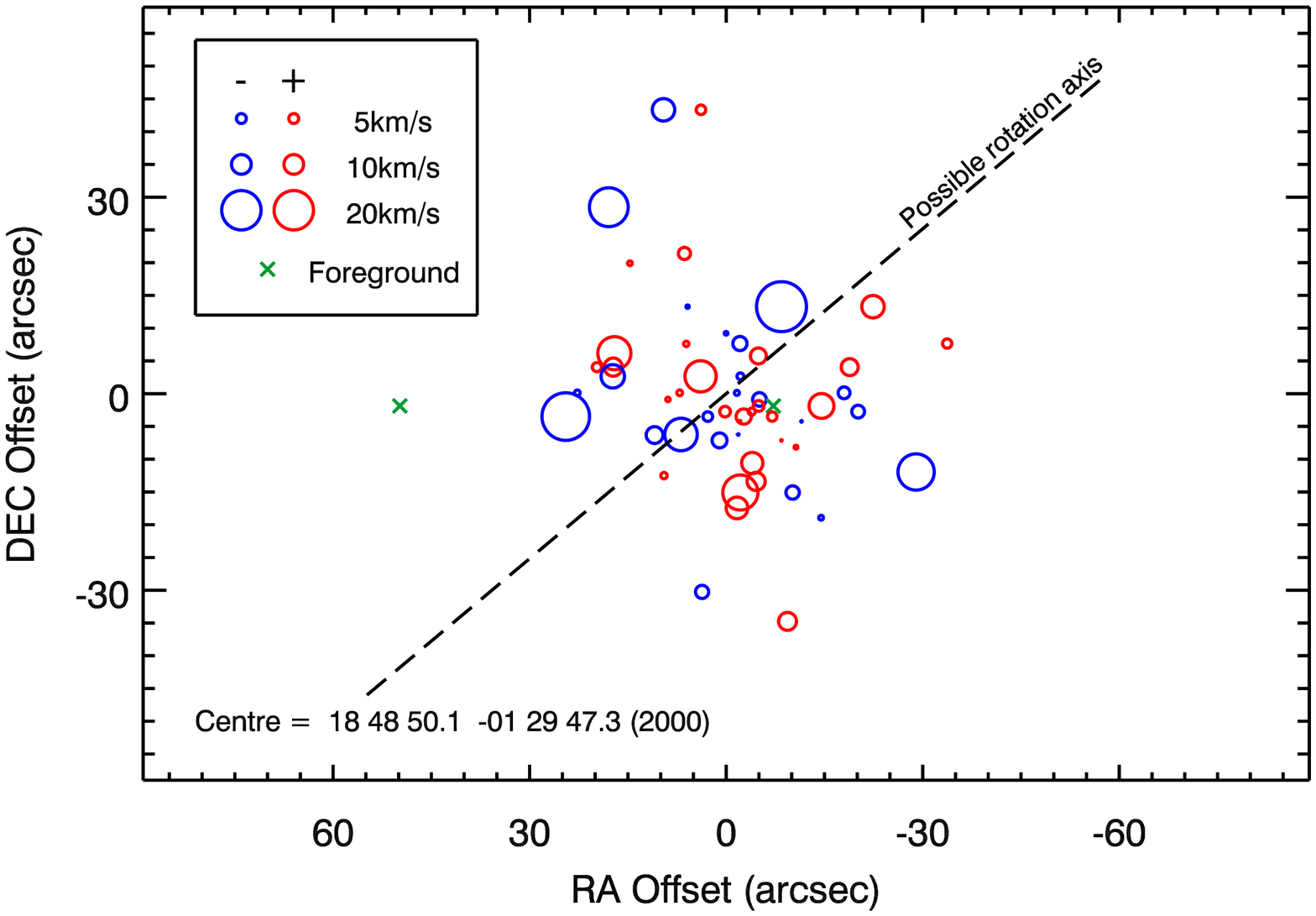}
  \caption{{\it Left}: A $K$-band wide-field image of the cluster taken from
    UKIDSS, with the stars for which we obtained spectroscopy
    identified by circles. {\it Right}: An illustration of the
    cluster's velocity dispersion. Symbol sizes denote the magnitude
    of the stellar velocities, once the average cluster velocity has
    been subtracted. Objects with relative velocities greater than
    30\kms\ were deemed to be unrelated foreground objects, and are
    identified by crosses.  }
  \label{fig:velpos}
\end{figure*}

We employed the observing strategy of stepping the slit across the
cluster to positions aligned with bright stars. This technique also
obtains data on fainter stars which fall into the slit serendipitously
at each position. The observations were split across 2 observing
blocks (OBs): one which included the seven slit positions covering the
brightest stars, with integration times of 60s (NDIT=1); and the
remaining 30 slit positions which had 90s (NDIT=1) integrations. At
the end of each OB arc lamp observations were taken for wavelength
calibration purposes. We observed the B5{\sc v} star Hip094378 to
measure the telluric absorption immediately after the science
observations. Flat-field and dark frames were taken at the end of the
night, while at the beginning of the night a bright standard star was
observed at several stepped positions along the slit to characterize
any spatial distortion on the detector.

Our data reduction began with subtraction of the dark current and
dividing though by a normalized flat field. The spatial distortion of
the 2-D frames was determined from the stepped standard star
observations, while the distortion in the dispersion direction was
measured from the arc-lamp observations of each observing block. The
distortion in the 2-D frames was then corrected by resampling each
frame onto an orthogonal grid, resulting in an absolute wavelength
calibration accurate to $\pm$2\kms (as measured from the residuals
between the observed and literature values of the arc lines). The sky
emission in each 2-D frame was characterized by measuring the
background in regions containing no star traces and interpolating
across the full length of the slit. Stellar spectra were extracted by
summing across the rows at each channel, being careful to avoid
contamination by neighbouring stars in crowded regions. Telluric
absorption was removed by dividing through by the standard star
spectrum, after first correcting for the star's intrinsic continuum
slope by dividing through by a black-body curve.

In the reduced spectra it was apparent that there was a degree
of fringing of up to 10\% of the continuum, towards the red end of
each spectrum, with a period of $\sim$0.04\microns. To correct for
this we identified a spectrum in each observing block that had no
detectable spectral features (i.e. no CO bandhead) and had high
signal-to-noise. These stars are likely to be early $\sim$G-type
stars. This spectrum was then smoothed with a narrow filter to remove
noise and intrinsic features, and was then used as a measure of the
fringing pattern. The fringing pattern of each target star is a
function of the star's position on the slit, so the `fringe' spectrum
was first cross-correlated with each science spectrum before dividing
through. In \fig{fig:showspec} we plot four example spectra, showing the
quality of our data. We show bright and faint stars, as well as stars
with both weak and strong CO absorption. The spectra prior to the
fringing correction are also shown to illustrate that the magnitude of
this artifact is only minor.

To measure the radial velocity of each star observed, we
cross-correlated each spectrum with that of Arcturus
\citep[from][]{W-H96arct}, after first degrading the Arcturus spectrum
to match the spectral resolution of our observations. This
cross-correlation procedure provides relative velocities accurate to a
few one-hundredths of a pixel, or $<$1\kms. At this level we are
likely dominated by systematic errors, so for a conservative estimate
of the uncertainty on each velocity measurement we use the absolute
error in the wavelength calibration solution, $\pm$2\kms. Comparing
the radial velocities of the stars obtained from each of the two
observing blocks, we find that the two averages are within
$\pm$1.5\kms\ of one another, indicative of the absolute error on the
wavelength calibration.

\begin{figure}
  \centering
  \includegraphics[width=8cm]{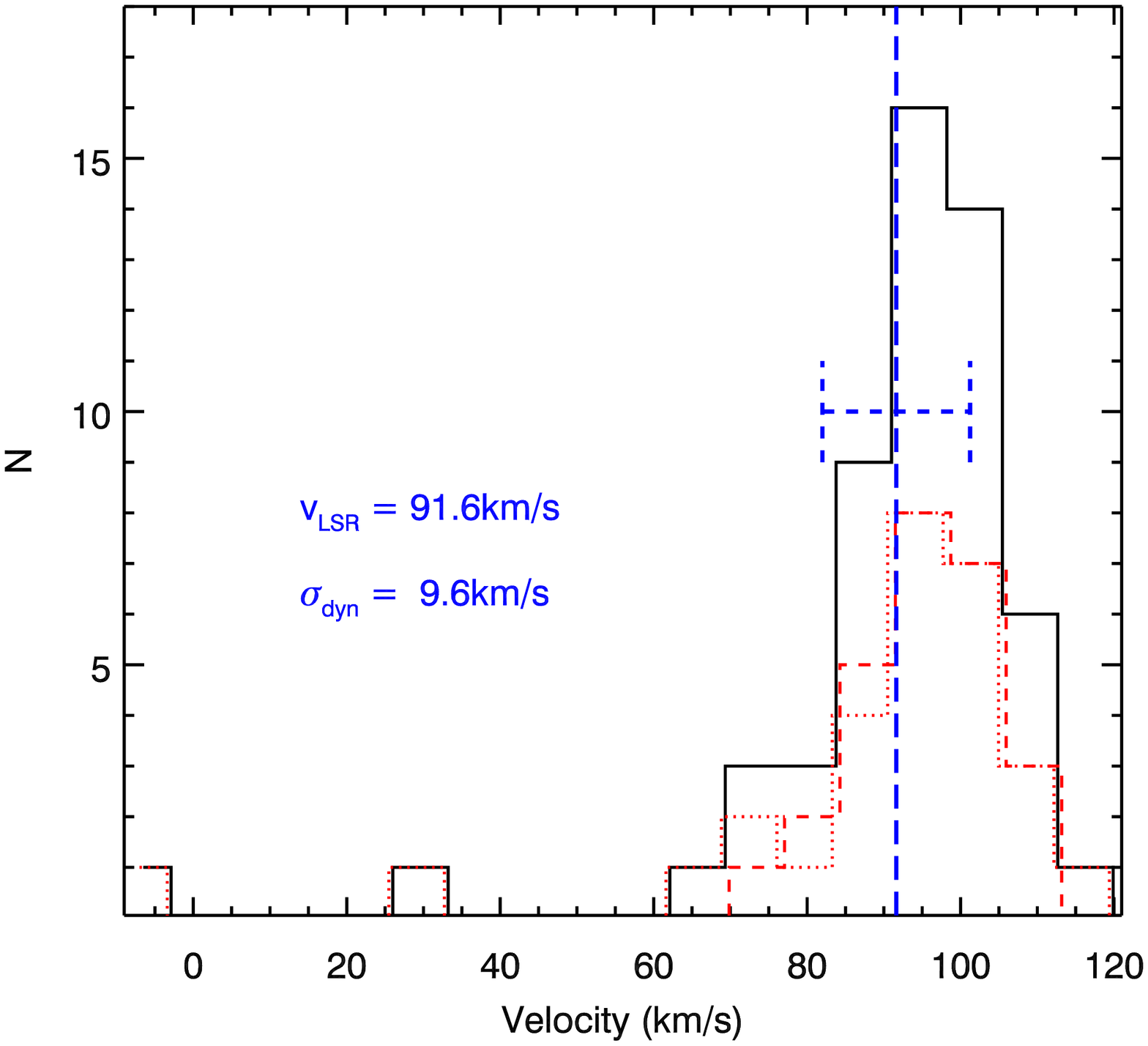}
  \caption{Histogram of the observed velocities. The mean velocity and
  its standard deviation (clipped at 92$\pm$30\kms to preclude foreground
  objects) is illustrated by the blue long-dashed line. The red dotted and
  dashed lines show the velocity distributions of the two separate
  observing blocks in which the data was taken. }
  \label{fig:velhist}
\end{figure}

\begin{table*}
  \centering
  \caption{Observational data of the stars observed. Astrometry is
    taken from slit positions, and is accurate to
    $\pm$1\arcsec. EW(CO) was measured between 2.293\microns\ and
    2.310\microns, uncertainties are $\pm$5\%, estimated from repeated
    measurements. Errors on velocities are conservatively $\pm$2\kms,
    dominated by the uncertainty in the wavelength
    calibration. Photometry is taken from \citet{Ivanov05}.  }
  \label{tab:results}
  \begin{tabular}{lccccccc}
    \hline
    ID & RA  & DEC & J & H & K & EW(CO) & $v_{\rm LSR}$ \\
       & (J2000) & (J2000) &   &   &   & (\AA) & (\kms) \\
    \hline \hline
 1 & 18 48 50.11 & -1 29 50.0 & 11.11 $\pm$ 0.06 &  8.89 $\pm$ 0.07 &  7.74 $\pm$ 0.06 &  28.3 &  97.0 \\
 2 & 18 48 49.77 & -1 29 41.6 & 11.85 $\pm$ 0.03 &  9.32 $\pm$ 0.04 &  8.00 $\pm$ 0.03 &  29.9 &  99.7 \\
 3 & 18 48 49.39 & -1 29 55.5 & 11.06 $\pm$ 0.12 &  9.53 $\pm$ 0.14 &  8.31 $\pm$ 0.11 &  34.0 &  93.6 \\
 4 & 18 48 50.53 & -1 29 25.9 & 12.85 $\pm$ 0.02 & 10.15 $\pm$ 0.02 &  8.86 $\pm$ 0.02 &  29.2 &  97.8 \\
 5 & 18 48 51.08 & -1 29 27.4 & 12.46 $\pm$ 0.02 & 10.04 $\pm$ 0.02 &  8.91 $\pm$ 0.02 &  26.2 &  94.0 \\
 6 & 18 48 51.24 & -1 29 41.1 & 12.65 $\pm$ 0.02 & 10.12 $\pm$ 0.02 &  8.95 $\pm$ 0.02 &  27.4 & 108.4 \\
 7 & 18 48 49.42 & -1 30  2.4 & 12.50 $\pm$ 0.02 & 10.10 $\pm$ 0.02 &  8.97 $\pm$ 0.01 &  26.1 &  84.8 \\
 8 & 18 48 49.54 & -1 29 34.0 & 10.54 $\pm$ 0.01 &  9.50 $\pm$ 0.01 &  9.00 $\pm$ 0.01 &  19.0 &  66.2 \\
 9 & 18 48 51.25 & -1 29 43.3 & 12.79 $\pm$ 0.02 & 10.28 $\pm$ 0.03 &  9.08 $\pm$ 0.02 &  28.0 & 100.8 \\
10 & 18 48 53.42 & -1 29 49.2 & 11.30 $\pm$ 0.01 &  9.92 $\pm$ 0.02 &  9.28 $\pm$ 0.01 &  23.5 &  -4.2 \\
11 & 18 48 48.84 & -1 29 43.3 & 13.38 $\pm$ 0.02 & 10.61 $\pm$ 0.02 &  9.39 $\pm$ 0.02 &   6.9 & 100.2 \\
12 & 18 48 48.17 & -1 29 59.3 & 15.40 $\pm$ 0.03 & 11.31 $\pm$ 0.03 &  9.53 $\pm$ 0.02 &  25.3 &  73.2 \\
13 & 18 48 51.62 & -1 29 47.2 & 13.10 $\pm$ 0.03 & 10.73 $\pm$ 0.03 &  9.65 $\pm$ 0.02 &  24.8 &  88.8 \\
14 & 18 48 50.36 & -1 29 44.7 & 13.01 $\pm$ 0.02 & 10.69 $\pm$ 0.02 &  9.68 $\pm$ 0.02 &   6.2 & 107.5 \\
15 & 18 48 49.84 & -1 29 57.9 & 13.07 $\pm$ 0.02 & 10.78 $\pm$ 0.02 &  9.72 $\pm$ 0.02 &  18.0 & 102.4 \\
16 & 18 48 49.80 & -1 30  0.7 & 13.14 $\pm$ 0.02 & 10.85 $\pm$ 0.03 &  9.78 $\pm$ 0.02 &  21.9 & 100.8 \\
17 & 18 48 50.36 & -1 29  4.0 & 13.54 $\pm$ 0.02 & 11.14 $\pm$ 0.02 &  9.92 $\pm$ 0.02 &   6.7 &  96.4 \\
18 & 18 48 49.48 & -1 30 22.1 & 13.35 $\pm$ 0.03 & 11.01 $\pm$ 0.03 &  9.95 $\pm$ 0.02 &  24.8 & 100.7 \\
19 & 18 48 50.83 & -1 29 53.6 & 13.16 $\pm$ 0.03 & 10.91 $\pm$ 0.03 &  9.96 $\pm$ 0.02 &   5.1 &  83.1 \\
20 & 18 48 48.76 & -1 29 50.0 & 13.82 $\pm$ 0.03 & 11.16 $\pm$ 0.03 &  9.98 $\pm$ 0.02 &  20.2 &  85.2 \\
21 & 18 48 49.13 & -1 29 49.2 & 14.63 $\pm$ 0.02 & 11.51 $\pm$ 0.03 & 10.06 $\pm$ 0.02 &  22.4 & 104.2 \\
22 & 18 48 50.17 & -1 29 54.4 & 13.82 $\pm$ 0.03 & 11.22 $\pm$ 0.03 & 10.07 $\pm$ 0.02 &  18.5 &  84.1 \\
23 & 18 48 49.13 & -1 30  6.2 & 13.38 $\pm$ 0.02 & 11.07 $\pm$ 0.03 & 10.07 $\pm$ 0.02 &   5.8 &  89.2 \\
24 & 18 48 50.10 & -1 29 38.1 & 13.36 $\pm$ 0.02 & 11.10 $\pm$ 0.03 & 10.10 $\pm$ 0.02 &   3.3 &  89.8 \\
25 & 18 48 51.73 & -1 29 50.8 & 13.14 $\pm$ 0.02 & 11.00 $\pm$ 0.03 & 10.10 $\pm$ 0.02 &   5.8 &  67.4 \\
26 & 18 48 49.63 & -1 29 50.8 & 13.33 $\pm$ 0.02 & 11.21 $\pm$ 0.03 & 10.14 $\pm$ 0.02 &   5.2 &  96.4 \\
27 & 18 48 51.30 & -1 29 18.8 & 13.56 $\pm$ 0.02 & 11.19 $\pm$ 0.03 & 10.15 $\pm$ 0.02 &   1.0 &  71.9 \\
28 & 18 48 50.35 & -1 30 17.6 & 13.51 $\pm$ 0.02 & 11.24 $\pm$ 0.03 & 10.26 $\pm$ 0.02 &   5.7 &  84.9 \\
29 & 18 48 50.74 & -1 29  4.0 & 14.36 $\pm$ 0.03 & 11.55 $\pm$ 0.03 & 10.31 $\pm$ 0.02 &  10.4 &  80.2 \\
30 & 18 48 49.54 & -1 29 54.4 & 13.65 $\pm$ 0.03 & 11.34 $\pm$ 0.03 & 10.34 $\pm$ 0.02 &   6.2 &  92.0 \\
31 & 18 48 50.56 & -1 29 53.5 & 13.77 $\pm$ 0.03 & 11.47 $\pm$ 0.03 & 10.38 $\pm$ 0.03 &  17.8 &  75.1 \\
32 & 18 48 50.29 & -1 29 50.8 & 13.77 $\pm$ 0.02 & 11.47 $\pm$ 0.02 & 10.43 $\pm$ 0.02 &  15.5 &  86.7 \\
33 & 18 48 49.96 & -1 29 44.7 & 14.00 $\pm$ 0.02 & 11.56 $\pm$ 0.02 & 10.45 $\pm$ 0.02 &  12.6 &  88.2 \\
34 & 18 48 50.51 & -1 29 39.7 & 13.88 $\pm$ 0.02 & 11.65 $\pm$ 0.03 & 10.62 $\pm$ 0.02 &  11.2 &  94.3 \\
35 & 18 48 49.99 & -1 29 47.2 & 13.83 $\pm$ 0.03 & 11.69 $\pm$ 0.03 & 10.62 $\pm$ 0.02 &   4.7 &  89.1 \\
36 & 18 48 47.85 & -1 29 39.7 & 14.42 $\pm$ 0.03 & 11.79 $\pm$ 0.03 & 10.65 $\pm$ 0.02 &   4.1 &  96.3 \\
37 & 18 48 50.49 & -1 29 34.0 & 13.98 $\pm$ 0.02 & 11.77 $\pm$ 0.02 & 10.76 $\pm$ 0.02 &  16.1 &  90.0 \\
38 & 18 48 49.96 & -1 30  2.4 & 14.10 $\pm$ 0.02 & 11.86 $\pm$ 0.03 & 10.87 $\pm$ 0.02 &  15.2 & 109.5 \\
39 & 18 48 51.25 & -1 29 44.7 & 14.17 $\pm$ 0.03 & 11.90 $\pm$ 0.04 & 10.87 $\pm$ 0.03 &  18.8 &  79.7 \\
40 & 18 48 49.99 & -1 30  4.7 & 14.36 $\pm$ 0.03 & 12.00 $\pm$ 0.03 & 10.94 $\pm$ 0.02 &  16.0 & 102.6 \\
41 & 18 48 49.98 & -1 29 53.5 & 14.20 $\pm$ 0.03 & 12.04 $\pm$ 0.03 & 10.96 $\pm$ 0.02 &   7.7 &  91.2 \\
42 & 18 48 49.76 & -1 29 48.2 & 14.18 $\pm$ 0.03 & 11.82 $\pm$ 0.03 & 10.98 $\pm$ 0.02 &  11.6 &  84.7 \\
43 & 18 48 49.84 & -1 29 50.0 & 14.81 $\pm$ 0.03 & 12.20 $\pm$ 0.03 & 11.04 $\pm$ 0.03 &  13.9 &  95.2 \\
44 & 18 48 49.92 & -1 29 50.8 & 14.46 $\pm$ 0.05 & 12.33 $\pm$ 0.06 & 11.09 $\pm$ 0.05 &   4.6 &  99.3 \\
45 & 18 48 48.90 & -1 29 47.2 & 15.25 $\pm$ 0.03 & 12.44 $\pm$ 0.03 & 11.13 $\pm$ 0.02 &  13.7 &  85.6 \\
46 & 18 48 49.96 & -1 29 51.5 & 14.46 $\pm$ 0.03 & 12.27 $\pm$ 0.04 & 11.19 $\pm$ 0.03 &   9.9 &  91.7 \\
47 & 18 48 50.57 & -1 29 47.2 & 14.62 $\pm$ 0.03 & 12.29 $\pm$ 0.04 & 11.21 $\pm$ 0.03 &  17.9 &  94.5 \\
48 & 18 48 50.73 & -1 29 59.8 & 14.25 $\pm$ 0.02 & 12.20 $\pm$ 0.03 & 11.24 $\pm$ 0.02 &  11.3 &  94.8 \\
49 & 18 48 51.42 & -1 29 43.3 & 14.61 $\pm$ 0.02 & 12.33 $\pm$ 0.02 & 11.34 $\pm$ 0.02 &   4.3 &  96.2 \\
50 & 18 48 50.69 & -1 29 48.2 & 14.40 $\pm$ 0.03 & 12.29 $\pm$ 0.03 & 11.34 $\pm$ 0.02 &   4.1 &  93.8 \\
51 & 18 48 48.61 & -1 29 34.0 & 15.36 $\pm$ 0.03 & 12.57 $\pm$ 0.03 & 11.39 $\pm$ 0.02 &   7.2 & 103.0 \\
52 & 18 48 49.62 & -1 29 49.2 & 13.07 $\pm$ 0.02 & 11.84 $\pm$ 0.02 & 11.49 $\pm$ 0.02 &   3.2 &  30.1 \\
53 & 18 48 49.33 & -1 29 51.5 & 15.25 $\pm$ 0.02 & 12.83 $\pm$ 0.02 & 11.61 $\pm$ 0.02 &  10.4 &  91.4 \\
54 & 18 48 49.77 & -1 29 49.2 & 14.90 $\pm$ 0.04 & 12.83 $\pm$ 0.04 & 11.71 $\pm$ 0.03 &   7.4 &  97.1 \\
55 & 18 48 49.96 & -1 29 39.7 & 15.12 $\pm$ 0.03 & 12.89 $\pm$ 0.03 & 12.13 $\pm$ 0.02 &  11.7 &  84.4 \\
\hline
  \end{tabular}
\end{table*}

\section{Results} \label{sec:results}

In Table \ref{tab:results} we list the observed properties for all
stars observed for which we could obtain corresponding near-IR
photometry. Astrometry for each star was determined from the telescope
pointing and the star's position on the slit, with UKIDSS images used
to fine-tune the absolute calibration of the RA and DEC in the file
headers. Photometry was obtained by cross-correlating with the
photometric catalogue of \citet{Ivanov05}. For six stars in our sample
no corresponding photometry could be found. These data were discarded
from the sample, since the signal-to-noise ratio (SNR) was much lower
than the rest, which typically have SNR$>$50. We checked for
systematic errors in our data by checking for correlations between
radial velocity and brightness, colour, slit position, coordinates and
CO absorption strength, finding no apparent trends.

\subsection{Radial velocities}
The stars observed are indicated in \fig{fig:velpos}, while in the
right-hand panel of the figure we illustrate their spatial velocity
distribution. A histogram of the observed velocities (in the Local
Standard of Rest frame) is shown in \fig{fig:velhist}. The majority of
the velocities are within 90$\pm$30\kms, with a small number of
measurements of much lower velocities. These objects with low
line-of-sight velocities are likely to be foreground objects which are
not physically related to the cluster. To find the average velocity of
the observed stars we took the mean of all measurements iteratively
clipped at 3$\sigma$ to discard objects with outlying velocities. We
find a mean velocity of 91.6\kms, with clipping limits $\pm$30\kms\ of
this value, and 53 stars having velocities falling within this
range. The formal error on the mean velocity ($\equiv
\sigma_{v}/\sqrt{n}$) is compounded by the absolute error on the
wavelength calibration, therefore our formal measurement of the
average cluster velocity is $91.6 \pm 2.4$\kms. Altering the clipping
limits by 10\kms\ in either direction does not affect the average
velocity outside this error margin.

To find the velocity dispersion, we calculate the r.m.s. standard
deviation on the sigma-clipped mean, and subtract the absolute error
in quadrature. Before making this calculation however, we first assume
that the stars observed in each OB are sampled from the same velocity
distribution, shifting each population by $\pm$1.8\kms\ so that they
have the same mean. Following this procedure, and clipping at
92$\pm$30\kms, we find $\sigma_{v} = 9.6$\kms. If we do {\it not
  shift} the velocities of the two OBs, we find $\sigma_{v} =
9.9$\kms. If we narrow the clipping limits by 5\kms\ we find
$\sigma_{v} = 9.0$\kms. Hence, our measurement is stable to within
$\pm0.6$\kms.

To estimate the formal empirical error on $\sigma_{v}$, we ran
monte-carlo experiments in which we generated a random set of
velocities with a gaussian distribution and $\sigma$=9.6\kms, and
folded in the experimental uncertainty of $\pm$2\kms. We then
determined what would be the observed velocity dispersion once the
instrumental resolution was removed. We ran these experiments for
several sizes of samples, from 10 to 100 measurements, and calculated
the empirical error for each. We found that for a sample size of 53
such as ours, the uncertainty on $\sigma_{v} = 9.6$\kms\ was
consistently 1.1\kms. Our formal measurement of GC01's velocity
dispersion is therefore $\sigma_{v} = 9.6 \pm 1.1$\kms.

\subsection{Cluster rotation}
In addition to the velocity dispersion, we also investigated the
cluster for evidence of rotation. The relaxation time for a cluster of
GC01's mass and size (see later) is $\sim 10^8$yrs, so it is unlikely
that any rotation would be retained if it had an age typical of
globular clusters, $\sim$10Gyr \citep{F-F85}. Instead, the presence of
rotation would suggest a much younger age.

We divided the cluster into two with a line going through the cluster
centre which had a position-angle $\theta$. We then considered the
mean velocities of the stars on opposite sides of this line as a
function of $\theta$. We found that the maximum velocity difference
was $\sim$2.5\kms\ at an angle of $\theta = 130 \pm 40$\degr\ east of
north (indicated on the right panel of \fig{fig:velpos}). This is
consistent with being perpendicular to the cluster's apparent
elongation ($\sim$60\degr, see K05\footnote{The {\it Spitzer}/GLIMPSE
  images of GC01 indicate that the cluster suffers from differential
  reddening, which may give the appearance of an elongated
  morphology}), suggesting an oblate-spheroidal structure to the
cluster.

To test the significance of this result we performed a number of
trials with randomized the velocities, keeping the stellar positions
fixed and repeating the same test for rotation. We found that our
result of a velocity gradient is significant at the 1.8$\sigma$
level. Hence, any evidence we see for rotation is marginal. We also
note that any rotational velocity contribution appears to be much
smaller than the overall velocity dispersion.

%MENTION ANALYSIS OF ROTATION -- METHOD OF ANALYSIS, AXIS OF ROTATION,
%SIGNIFICANCE, SHOW THAT ANY ROTATION IS MUCH SMALLER THAN V-DISP. 

%\begin{figure*}
% \includegraphics[width=8.5cm]{modplot_20.eps}
%  \includegraphics[width=8.5cm]{modplot_0.eps}
%  \caption{}
%  \label{fig:modplot}
%\end{figure*}

\begin{figure}
  \includegraphics[width=8.5cm]{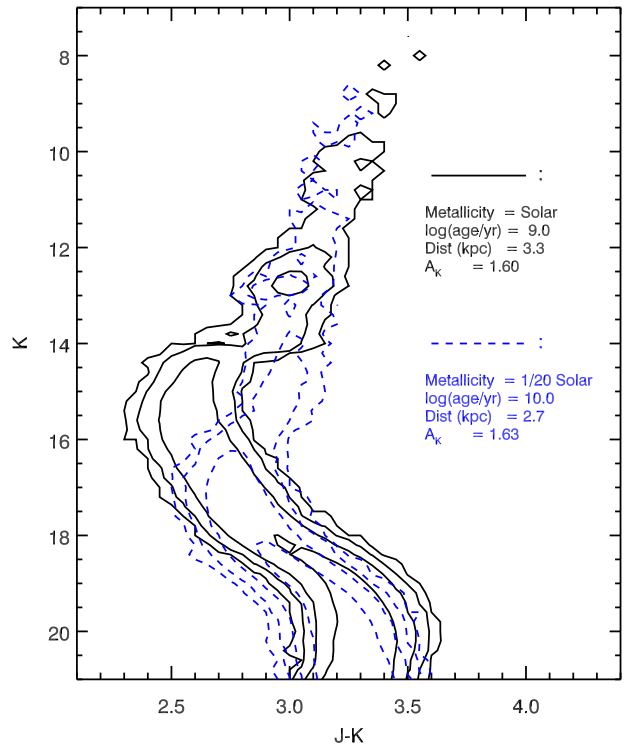}
  \caption{Hess diagram for two synthesized clusters, constructed
    using the evolutionary models of \citet{Cioni06a,Cioni06b}. The
    plot shows a metal-rich 1Gyr-old cluster (black solid contours)
    and a metal-poor, 10Gyr-old cluster (blue dashed contours). In the
    regime of current observations ($K<$14), the clusters are
    indistinguishable. Both have stellar overdensities at $K$=13, and
    have RGB slopes that may appear similar if the upper RGB is poorly
    populated. }
  \label{fig:hess}
\end{figure}

\begin{figure}
  \includegraphics[width=8.5cm]{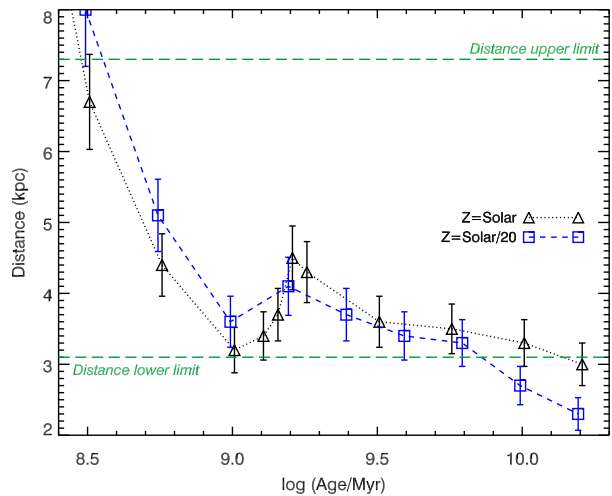}
  \caption{The distance to GC01, inferred from the IR brightness and
    colour of the Red Clump, depending on the cluster's age, using the
    evolutionary models of \citet{Cioni06a,Cioni06b}. The upper and
    lower limits to the cluster's distance, argued by K05 on
    the basis of the $^{13}$CO column density, are marked by the green
    dashed lines. }
  \label{fig:age_v_dist}
\end{figure}

\subsection{Distance} \label{sec:dist}
K05 use the $^{13}$CO integrated column density along the
line-of-sight to GC01 to constrain the distance to the cluster. They
argue that the extinction to the cluster is consistent with being
behind a $^{13}$CO cloud at \vlsr=46\kms, but {\it in front} of clouds
at 81\kms\ and 100\kms. As the latter velocity corresponds to the
tangent point at 7.3kpc, while the \vlsr=46\kms\ cloud has a kinematic
distance of 3.1kpc, K05 conclude that the distance to GC01 must be
3.1-7.3kpc. 

If, for the moment, we assume that the cluster is part of the disk
rather than a globular passing through the disk, then the measured
velocity \vlsr=90$\pm$4\kms\ yields kinematic distances of $d_{\rm
  near} = 5.0\pm0.9$kpc and $d_{\rm far} = 7.9\pm0.4$kpc (using the
rotation curve of \citep{B-B93}; see \citet{RSGC2paper} for more
details). If we discard the far-side velocity on the basis that the
cluster is unlikely to be beyond the tangent point (K05), then we find
a consistent picture whereby the cluster is located at 5.0$\pm0.9$kpc,
behind the molecular cloud at 3.1kpc which is responsible for the
majority of the line-of-sight extinction. Thus, GC01 radial velocity
is consistent with the object co-moving with the disk.

\begin{figure*}
  \centering
  \includegraphics[width=8.75cm,bb=94 360 762 906]{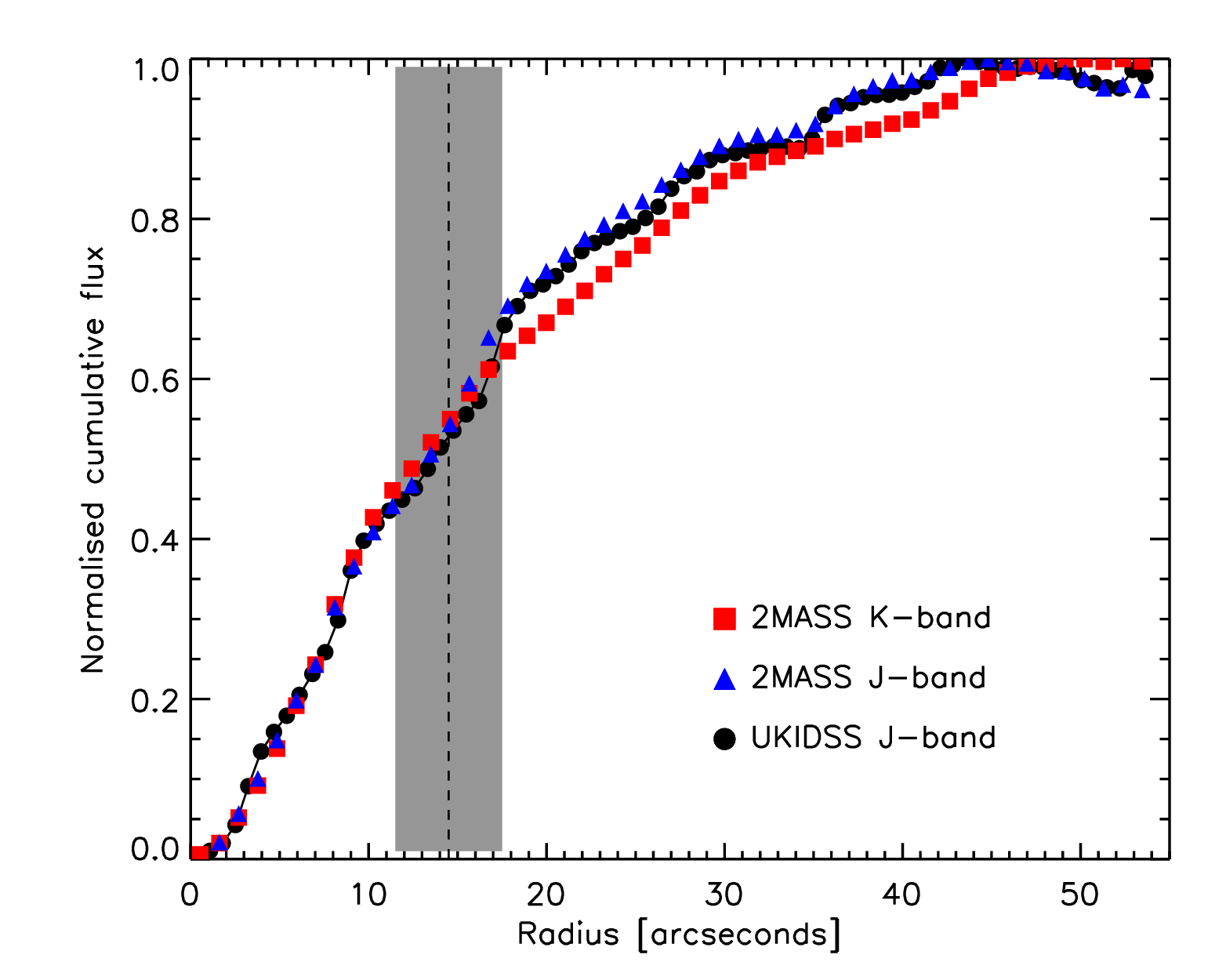}
  \includegraphics[width=8cm,bb=40 10 603 518]{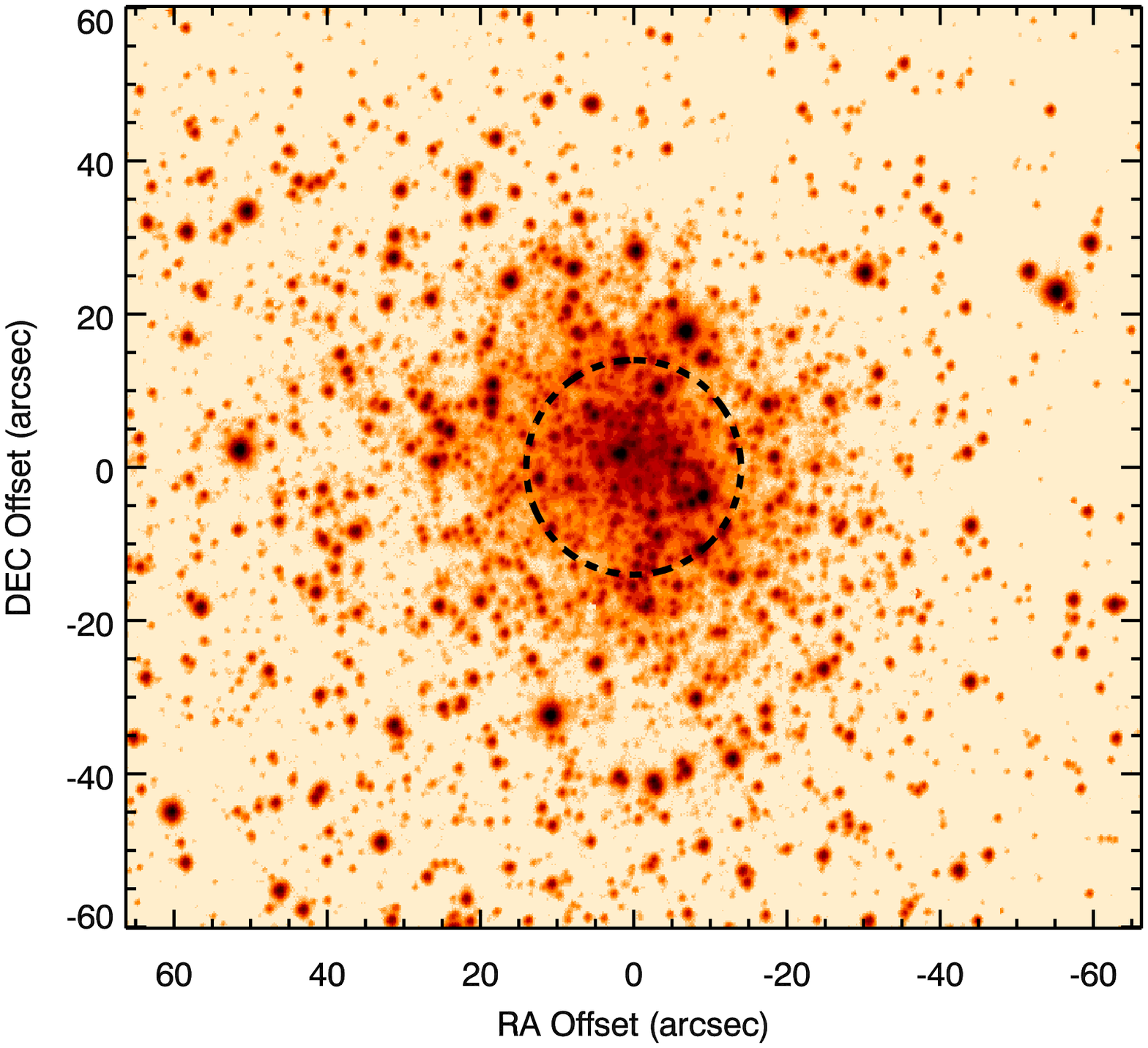}
  \caption{{\it Left}: Surface-brightness profile of the cluster from
    observations at different telescopes and through different
    filters, all of which give very similar values for $R_{\rm eff}$
    (indicated by the dashed line). The shaded area indicates the
    magnitude of the uncertainty on $R_{\rm eff}$.  {\it Right}:
    UKIDSS $K$-band image of the cluster, logarithmically scaled
    between the background level and 1000$\sigma$ above the
    background, with the half-light radius illustrated by the dashed
    circle. }
  \label{fig:radius}
\end{figure*}

In another study of the cluster's distance, \citet{Ivanov05} detected
what appears to be the Red Clump in a near-IR photometric study of
GC01, located at $K \approx$13 and $J-K \approx$3. These authors used
this feature to estimate the distance to the cluster. However, they
only used a single calibration for the absolute magnitude of the Red
Clump, when in fact the location of this feature depends upon cluster
age and metallicity \citep{Pietrzynski03}. Here, we have re-evaluated
the distance to GC01 using the Red Clump method. Taking the
evolutionary models of \citet[][ and references
  therein]{Cioni06a,Cioni06b} and a Kroupa IMF \citep{Kroupa02}, we
generated synthetic populations of stars for a range of ages and in
two distinct metallicity regimes: Solar (i.e. metal-rich) and 1/20th
Solar (i.e. metal-poor). For each synthetic cluster we created a $K$
vs $J-K$ colour-magnitude diagram (CMD), and adjusted the cluster's
extinction and distance until a good match to the simulated and
observed Red Clump was found. We show two such simulations in
\fig{fig:hess}, where it can be seen that clusters with very different
ages and metallicities can produce stellar overdensities at the same
location in the CMD.

The results of this experiment are shown in \fig{fig:age_v_dist}. The
age and metallicity dependence on the Red Clump distance calibration
is rather weak for clusters older than $\sim$500Myr, while for
clusters younger than this the relation is very sensitive to age. We
see that, within the distance upper and lower limits argued by K05,
values of $\log$(age/Myr) = 8.5 -- 10 (300Myr -- 10Gyr) are permitted
at either metallicity. The implied extinction is also dependent on the
cluster age, ranging from $A_{K}=1.52-1.61$ in the metal-rich models
and $A_{K}=1.63-1.73$ in the metal-poor models.

To summarize, we have insufficient evidence to point towards a
definitive cluster distance within the upper and lower limits of
3.1--7.3kpc. If the cluster is comoving with the Galactic disk, as
suggested by its radial velocity, then the cluster is unlikely to be
older than $\sim$1Gyr, and so we have a consistent picture from the
Red Clump and kinematic distances in which the cluster is located at
$\approx$5kpc. However, if the cluster is an old globular, with a low
metallicity and an age $\ga$10Gyr, then the location of the Red Clump
suggests that the cluster distance is close to the lower limit
(3--3.5kpc), consistent with the distance determination of
\citet{Ivanov05}, 3.7$\pm$0.8kpc. In this case the cluster would
likely be traversing the Galactic disk, and the similarity between the
cluster's kinematic and physical distances would be purely
coincidental. Comparing the observed radial velocity with the
distribution of old globular clusters, we find that there is a 10\%
chance that it has a velocity that happens to match the expected
galactic rotation, calculated using the velocity distribution of
globular clusters in \citet{Harris96}. Hence, though the radial
velocity suggests that GC01 is part of the disk population, this
evidence alone cannot provide definitive evidence one way or the other
to the cluster's nature.

\subsection{Cluster size}
In order to estimate the effective radius $R_{\rm eff}$ of the cluster
we utilise imaging from the 2MASS (J, H, and Ks) \citep{Skrutskie06}
and UKIDSS (we only used the J-band images as the H and K-band images
suffered severe saturation effects) near-IR surveys.  In both cases we
estimated the position of the cluster center using isophotes at
distances larger than 10\arcsec\ from the determined center.  We then
summed the flux in concentric circles, centered on this point, and
subtracted the (area normalised) flux in a background annulus
$>40$\arcsec\ from the cluster center.  This cumulative flux
distribution was then normalized to the maximum value attained, and
the effective radius was defined as the radius containing half the
light of the cluster. Our measurement of the cluster's flux profile is
shown in the left panel of \fig{fig:radius}.

In order to test the stability and estimate the error of our effective
radius determination, we carried out the same analysis adopting
different cluster centres in a box of width and height of
8\arcsec\ centred on the original position.  Additionally, we varied
the background annulus from 40\arcsec\ to 90\arcsec.  The mean of
these tests was $R_{\rm eff}$=14.5\arcsec. This is shown as a dashed
circle in the right panel of \fig{fig:radius}.  The error was then
estimated as the standard deviation of all of these trials, and found
to be 3\arcsec. Overall, the agreement between the derived effective
radius using the 2MASS and UKIDSS images was very good.

This radius is significantly smaller than that reported by K05
(36\arcsec). As an extra check on our radius measurement, we repeated
our analysis on the GLIMPSE images of the cluster, finding
measurements consistent with that from near-IR images
(14.1$\pm$0.8\arcsec\ and 13.8$\pm$0.8\arcsec\ for the
3.6\microns\ and 4.5\microns\ images respectively). It is not clear
where the difference between our measurement and that of K05 stems
from, but one possibility lies in the estimate of the (high)
background, due to the location of the cluster in the Galactic plane.

Assuming a distance to the cluster of 3.1, 5.0 or 7.3~kpc (see
Section~3.1), results in an effective radius of 0.22, 0.35 and
0.51~pc, respectively.  Hence, G01 is a very dense cluster and appears
to be much more compact than typical globular clusters
\citep{Harris96} and young/intermediate aged massive clusters in
external galaxies \citep{Scheepmaker07}.

\subsection{Cluster mass}
We now calculate the mass of GC01 under the assumption that the
cluster is in virial equilibrium. It has recently been argued that for
very young clusters this assumption may not be valid, as the removal
of intra-cluster gas by supernovae leaves the velocity dispersion of
the cluster super-virial. However, this effect is thought to be
important only for clusters with ages $<$20Myr
\citep{B-G06,G-B06}. Binaries can also influence the velocity
dispersion of a stellar cluster, though this effect becomes less
important for clusters older than $\sim$100Myr, especially if the
cluster is massive and compact \citep{Gieles10}. We neglect any
contribution from binary motions in our analysis of the velocity
dispersion.

To determine the dynamical mass $M_{\rm dyn}$ we use the equation,

\begin{equation}
  M_{\rm dyn} = \eta \frac{R_{\rm eff} \sigma_{\rm v}^{2}}{G}
  \label{equ:mdyn}
\end{equation}

\noindent where $R_{\rm eff}$ is the cluster radius, $G$ is the
gravitational constant, and $\eta$ depends on the cluster density
distribution but is typically taken to be $\approx$10. Using the value
of $R_{\rm cl} = 14$\arcsec\ measured earlier, we find a cluster mass
of $(8\pm3) \times 10^{4}$\msun, where the error is dominated by the
error in distance. While this is a typical mass for a globular
cluster, young clusters in this mass range have also been observed in
the Galactic plane \citep[e.g. Wd~1,
  RSGC2][]{Brandner08,RSGC2paper}. Indeed, for a disk-borne cluster to
survive until an age of $\sim$1Gyr without being dissolved it would
likely need to be at least this massive \citep{Lamers05}.

GC01 appears to be an extremely dense and compact cluster, with a
density of $\log (\rho_h / M_{\odot}\,pc^{-3} ) = 5.0\pm0.4$ (where
$\rho_h = 3M_{\rm cl} / (8\pi R_{\rm h}^3)$, and the half-mass radius
$R_h \equiv 4/3 R_{\rm eff}$). Here, the uncertainty is again
dominated by that on the cluster's distance. This is density exceeds
that any globular cluster in the catalogue of \citet{Harris96}; the
most dense is NGC~6540 with $\log (\rho_h) = 4.9$, while all globular
clusters with masses similar to GC01 have densities an order of
magnitude lower \footnote{In calculating $\rho_{\rm h}$ we have
  assumed that the ratio of mass to $V$-band luminosity $M/L_{V}
  \approx 2$ \citep{McL-vdM05}.}. The density is however comparable to
very young massive clusters, such as the Arches \citep{Kim06}.
Recently, \citet{Gieles10exp} have performed analytic and N-body
calculations to look at the evolution of stellar clusters, in
particular their size and density. These authors find that clusters
cannot exist for indefinite periods in extremely dense state, but
rather expand on the timescales of a few relaxation times.  If GC01 is
an old globular, we would have expected it to have expanded during the
past 10-12 Gyr, and hence the current dense state of GC01 argues for a
much younger ($\la$1 Gyr) age.

\section{Discussion: the nature of GLIMPSE-CO1} \label{sec:disc}
We now address the topic of GC01's nature -- specifically, is it an
old, metal-poor globular cluster passing through the plane of the
Galactic disk, or is it a young, metal-rich disk-borne cluster?

The discussion in K05 provides convincing arguments against the
cluster being very young, i.e. $\la$50Myr, due to the lack of bright
Red Supergiants, OB stars and intra-cluster material. However, as we
have already discussed in Sect.\ \ref{sec:intro}, their analysis is
unable to distinguish conclusively between the cases of a very old
metal-poor (i.e. globular) cluster and that of a metal-rich
intermediate age ($\sim$1Gyr) cluster.

We also mention the detection of extended X-ray emission around GC01 by
\citet{Mirabal10}, spatially coincident with the mid-IR emission seen
by K05. \citet{Mirabal10} speculate that this emission may be arising
in a bow-shock structure, created as the cluster plunges through the
disk, although they note that there are other explanations for the
emission which cannot be ruled out.

In this paper we have shown from GC01's velocity dispersion that the
cluster's mass is $(8 \pm 3) \times$10$^4$\msun. This mass is typical
for globular clusters \citep[see e.g.][]{Harris99}, while it is
certainly not unheard of for young Galactic clusters either. Hence,
the cluster mass does not provide a persuasive argument for either
side of the debate.

Similarly, the cluster's radial velocity does not rule out either
argument. A radial velocity inconsistent with the local Galactic
rotational velocity would have been a strong indication that the
cluster was not co-moving with the disk, and so was likely an old
globular passing through. However, the velocity we measure {\it is}
consistent with the Galactic rotation curve at a distance of
$\sim$5kpc, which is perfectly within the upper and lower distance
limits of 3-7kpc provided by K05. Though suggestive of a disk-borne
cluster, we find a non-negligible probability (10\%) that a globular
cluster passing through the disk would have this velocity. 

Our analysis of the Red Clump detected by \citet{Ivanov05} shows again
that, within the accepted distance range, the cluster could have an
age between several $\times$100Myr and 10Gyr. However, ages older than
10Gyr appear to be ruled out as this would place the cluster {\it in
  front} of the nearby molecular cloud, incompatible with the
cluster's extincition. This is one argument against the cluster being
an old globular. In addition, the cluster's high density, as well as
marginal evidence for rotation, are both suggestive of an age much
younger that that typical of globular clusters.

We attempted to measure the RGB slope for a sample of synthesized
clusters with a range of ages and metallicities, in order to compare
to the results of \citet{Ivanov05}. However, large uncertainties in
the measured slope caused by the Red Clump stars prevented us from
reaching any firm conclusions, since the observations were consistent
with both scenarios for the cluster's age and metallicity.

\begin{figure}
  \centering
  \includegraphics[width=8.5cm]{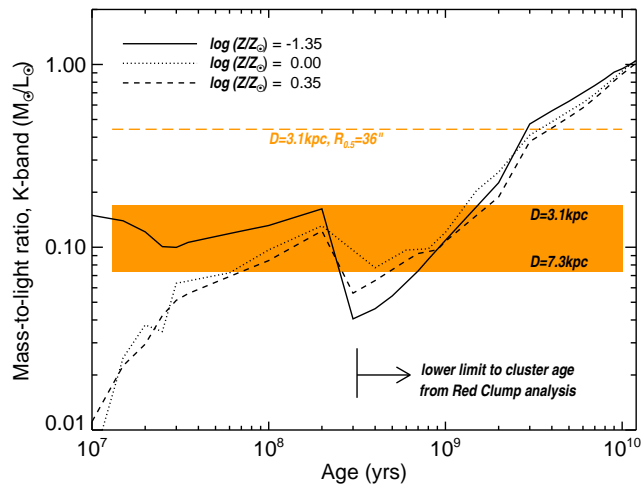}
  \caption{The $K$-band mass-to-light ($M$/$L$) ratio of a cluster as
    a function of age for three different metallicities, using the
    calculations of \citet{M-T00} and \citet{Maraston03}. As the
    $M$/$L$ we measure for GC01 depends on the cluster's physical
    size, which in turn depends on the distance, we indicate the
    cluster's full possible $M$/$L$ range with the shaded area. The
    maximum $M$/$L$ when using K05's measure of the cluster size is
    indicated by the long-dashed line. The arrow indicates the lower
    limit to the cluster age from the Red Clump analysis in
    Sect.\ \ref{sec:dist}.  }
  \label{fig:mtol}
\end{figure}

%\subsection{Evidence from the cluster's $M/L$ ratio} \label{sec:mtol}

In order to examine the nature of GC01 further, in \fig{fig:mtol} we
compare the cluster's mass-to-light ratio ($M/L$) to predictions made
by the stellar evolutionary models of \citet{M-T00} and
\citet{Maraston03}. The figure shows how $M/L$ varies in a cluster with
a Kroupa IMF as a function of age at different metallicities. Our mass
measurement of GC01 depends upon the cluster's size, and so depends
linearly on the cluster's distance $d$; while the cluster's luminosity
goes as $d^2$. Therefore, $M/L \propto 1/d$. For $L$, we use the
integrated $K$-band luminosity of K05, which we checked by
re-measuring from both 2MASS and UKIDSS images, being careful to treat
properly the effects of saturated stars.

In \fig{fig:mtol} we indicate the observed range of $M/L$ for the
upper and lower distance limits. We see that, for ages greater than
$\sim$2Gyr, there begins to be a large disagreement between the
observed $M/L$ and the model predictions. Even using the larger
cluster size as measured by K05 (indicated by the dashed line in
\fig{fig:mtol}) the maximum allowed age is 3Gyr. This is consistent
with our analysis of the Red Clump, which suggested an age between
300Myr -- 10Gyr (see Sect.\ \ref{sec:dist}). Therefore, we find a
cluster age of 0.3-2\,Gyr, which strongly indicates that GC01 is {\it
  not} an old globular cluster. This agrees with our findings of
extremely high stellar density and marginal evidence for rotation.

If, rather than being an old globular, the cluster is instead part of
the disk population, we can use take the kinematic distance of
5.0$\pm$0.9kpc and use \fig{fig:age_v_dist} to further constrain the
age of the cluster to between 0.4-0.8\,Gyr, which is perfectly
consistent with the $M/L$ ratio. We can then also constrain the mass
of the cluster to (8$\pm$2)$\times10^4$\msun. 
 
GC01 therefore represents the first detection of a very massive
($M>10^4$\msun) intermediate-age ($10^8-10^9$yr) cluster in our
Galaxy. Since star clusters are gradually disrupted over their
lifetimes \citep{Lamers05} it is likely that the initial mass of GC01
was much greater that indicated by its current velocity dispersion,
possibly by an order of magnitude. This would make it by far the most
massive cluster known to have formed in the last 10Gyr. Interestingly,
the age of the cluster roughly coincides with peaks in the
star-formation rate of the Solar neighbourhood and of both Magellanic
Clouds (MCs) approximately 400Myr ago \citep{Lamers05,H-Z04,H-Z09}. It
has been speculated that these starburst events were caused by an
interaction between the Galaxy and the MCs, and it is possible that
the formation of GC01 was triggered during this episode.

\subsection{Future studies}
There are some caveats to the $M/L$ analysis. Firstly, though the
cluster's $M/L$ is anomolously low for a globular, if it had passed
through the Galactic disk numerous times, and was mass segregated, one
may expect the lower mass stars at large distances from the cluster
core to have been stripped away, which would drive $M/L$
downward. Also, $M/L$ may be somewhat sensitive to the precise nature
of the IMF at sub-Solar masses \citep[for a review of the IMF in
  globular clusters, see][]{Bastian10}, and the $K$-band luminosities
of RGB and AGB stars. For these reasons, the nature of GC01 still open
for debate. Below, we suggest further experimental tests that should
provide incontrovertible evidence as to the nature of GC01.

\paragraph*{Metallicity studies:} \footnote{We note that, while
  \citet{Ivanov05} attempted to measure the metallicity of GC01 from
  the slope of the red giant brach, this technique is calibrated only
  for old globular clusters. Attempts are currently underway to
  generalize this relation for clusters with a range of ages, see
  \citet{Sharma10}.}  A cluster with an age typical of globular
clusters would be expected to have a metallicity well below Solar
\citep[$\log(Z/Z_{\odot})$ between -1.5 and -0.5, see][]{B-S06}. On
the other hand, an intermediate age cluster's metallicity would be
much closer to Solar. Analysis of the cluster members' spectra should
readily indicate which regime the cluster belongs to, since the
strengths of metallic lines in the infrared are very sensitive to
abundance levels \citep[e.g.][]{rsg_jband}.

\paragraph*{Deep photometry:} 
In \fig{fig:hess} we show that, with current photometric data, we are
unable to distinguish between the CMDs of the two regimes. However,
deeper photometry ($K>19$) would detect the location of the
main-sequence turn-off (MSTO) and it's position relative to the Red
Clump. Going deeper still ($K>20$), one is able to detect the kink in
the main-sequence caused by the onset of molecular hydrogen
absorption, which in conjunction with the MSTO has been shown to be a
powerful diagnostic of age, distance and metallicity
\citep{Bono10}. An analysis involving two-colour photometry would be
preferential to simple analysis of the cluster's luminosity function,
as differential extinction and field star contamination would make the
luminosity function difficult to interpret. From comparisons to
evolutionary predictions such as those in \fig{fig:hess}, deep IR
photometry should provide the best estimate of the cluster's age
and distance.

\section{Conclusions} \label{sec:conc}
Using high-resolution near-IR spectroscopy of over 50 stars in
GLIMPSE-C01, we have derived a dynamical mass for the cluster of
$(8\pm3) \times$10$^4$\msun. Using our observations in conjunction
with those in the literature, we have attempted to determine the
nature of the cluster: whether it is an old globular cluster passing
through the disk, or an intermediate age disk-borne cluster. The
cluster's radial velocity, which could have ruled out the intermediate
age possibility, instead indicates that the cluster is co-moving with
the disk. Our analysis of the cluster's Red Clump, the mass-to-light
ratio, as well as the marginal evidence we find of cluster rotation,
all indicate an age of $\la$ 1Gyr. From our results we conclude that
the cluster is part of the disk population, and we use the kinematic
distance to constrain the cluster's age to 400-800Myr. The cluster is
therefore the Galaxy's most massive intermediate age cluster
discovered to date. In addition to our observations, we have outlined
future observational investigations capable of providing a critical
test our hypothesis.

\section*{Acknowledgments}
We thank the referee Valentin Ivanov for his comments and suggestions
which helped us improve the paper, and for kindly supplying us with
the photometry of GLIMPSE-C01. We also thank Don Figer for useful
discussions. The data in this data was obtained through ESO under the
observing programme 383.D-0025. This work makes use of the UKIDSS
survey; the UKIDSS project is defined in \citet{Lawrence07}. UKIDSS
uses the UKIRT Wide Field Camera \citep[WFCAM][]{Casali07} and a
photometric system described in \citet{Hewitt06}. The pipeline
processing and science archive are described in and
\citet{Hambly08}. This publication makes use of data products from the
Two Micron All Sky Survey, which is a joint project of the University
of Massachusetts and the Infrared Processing and Analysis
Center/California Institute of Technology, funded by the National
Aeronautics and Space Administration and the National Science
Foundation.

\bibliographystyle{/fat/Data/bibtex/apj}
\bibliography{/fat/Data/bibtex/biblio}

%\appendix

\end{document}